\documentclass[pdflatex,sn-vancouver-num]{sn-jnl}

\usepackage{graphicx}%
\usepackage{multirow}%
\usepackage{amsmath,amssymb,amsfonts}%
\usepackage{amsthm}%
\usepackage{mathrsfs}%
\usepackage[title]{appendix}%
\usepackage{xcolor}%
\usepackage{textcomp}%
\usepackage{manyfoot}%
\usepackage{booktabs}%
\usepackage{comment}
\usepackage{algorithm}%
\usepackage{algorithmicx}%
\usepackage{algpseudocode}%
\usepackage{listings}%

\theoremstyle{thmstyleone}%
\theoremstyle{thmstyletwo}%

\theoremstyle{thmstylethree}%

\begin{document}

\title[Article Title]{Parameter Identifiability Under Limited Experimental Data in Age-Structured Models of the Cell Cycle}


\author*[1]{\fnm{Ruby E.} \sur{Nixson}}\email{ruby.nixson@maths.ox.ac.uk}

\author[1,2]{\fnm{Helen M.} \sur{Byrne}}
\equalcont{These authors contributed equally to this work.}

\author[3]{\fnm{Joe M.} \sur{Pitt-Francis}}
\equalcont{These authors contributed equally to this work.}

\author[1]{\fnm{Philip K.} \sur{Maini}}
\equalcont{These authors contributed equally to this work.}

\affil[1]{\orgdiv{Wolfson Centre for Mathematical Biology}, \orgname{University of Oxford}, \orgaddress{\city{Oxford}, \country{UK}}}

\affil[2]{\orgdiv{Ludwig Institute for Cancer Research}, \orgname{University of Oxford}, \orgaddress{\city{Oxford}, \country{UK}}}

\affil[3]{\orgdiv{Department of Computer Science}, \orgname{University of Oxford}, \orgaddress{\city{Oxford}, \country{UK}}}


\abstract{The mitotic cell cycle governs DNA replication and cell division. The effectiveness of radiotherapy and chemotherapy depends on cell-cycle position, with increased resistance during DNA replication and mitosis. Thus, accurate mathematical models of the cell cycle are essential for understanding and predicting treatment response. However, mathematical modellers often face the problem of a lack of publicly available, sufficiently resolved, time-series datasets for parametrising models. In this work, we consider how the ability to collate population summary measurements across the literature, from different cell lines and/or experimental set ups, affects identifiability of parameters for a cell cycle model.

Initially synchronised cell populations gradually desynchronise over successive cycles, converging to balanced exponential growth (BEG) which is characterised by exponential population growth and steady, time-independent phase proportions. These proportions can be obtained from fluorescence-activated cell sorting (FACS) data. The increasing use of the Fluorescent Ubiquitination-based Cell Cycle Indicator (FUCCI) provides higher-resolution information on phase dynamics, such as minimum phase durations and variability.

We present an age-structured PDE model in which cell-cycle phase progression follows a delayed gamma distribution. We derive analytical expressions for BEG phase proportions and other FUCCI-observable quantities, and use them to assess how data availability influences parameter identifiability. When parameters are not uniquely identifiable, we determine identifiable parameter groupings, thereby determining the minimum amount of data that must be available for successfully fitting structured population models of the cell cycle.}


\maketitle

\section{Introduction}\label{sec1}

The mitotic cell cycle governs the duplication of cellular content and division of a cell into two daughter cells. The cycle is comprised of four stages that must be completed sequentially before division can take place. Firstly, a cell enters the $G_1$ phase, in which the cell size grows and preparation for DNA replication begins. Within this phase, a cell faces a ``choice" to either remain in an actively cycling state, or to pause cell cycle progression in a reversible manner, known as quiescence or the $G_0$ phase ~\citep{matthewsCellCycleControl2022}. DNA replication takes place in the second phase, $S$, before progression into the $G_2$ phase, in which preparation for division occurs, as well as repair of any damage accrued during the $S$ phase. Following this, the cell enters mitosis, $M$, a short phase culminating in cell division.

One common treatment for cancer is radiotherapy, and experimental results show that response to radiotherapy can be affected by cell cycle phase~\citep{pawlikRoleCellCycle2004, lonatiRadiationinducedCellCycle2021, muzRoleHypoxiaCancer2015}. Cells in the M phase are most radio-sensitive, whilst cells in the S phase are most radio-resistant~\citep{yasharBasicPrinciplesGynecologic2018}. Quiescent cells are also radio-resistant~\citep{lonatiRadiationinducedCellCycle2021}. Furthermore, chemotherapies for cancerous cells are also known to interact with the cell cycle, with some chemotherapeutic drugs modulating progress through the cell cycle, and others inducing a cell cycle phase-dependent response~\citep{schwartzTargetingCellCycle2005, sunInfluenceCellCycle2021, ottoCellCycleProteins2017}. Therefore, accurate mathematical models of the cell cycle are essential when applying them to understanding and predicting treatment responses.

Traditionally, flow cytometry has been the most common technique used to report cell cycle phases in a population of cells. By fixing the cells and staining for DNA content, flow cytometry can distinguish between cells with one copy of each chromosome (indicating cells in $G_1/G_0$) from cells with two copies of each chromosome (post-replication, so $G_2$ and $M$ cells), and those with intermediate DNA content during replication in $S$ phase~\citep{pozarowskiAnalysisCellCycle2004, cecchiniAnalysisCellCycle2012, kimAssayingCellCycle2015, terryChapter16Cell2001}. Staining with propidium iodide (PI) indicates a cell's DNA content, but cannot accurately distinguish between cells in the late-$G_1$ and early-$S$ phases, or between cells in the late-$S$ and early-$G_2$ phases. 5-bromo-2'-deoxyuridine (BrdU) staining is then used, which identifies cells in the $S$ phase since it is incorporated into newly-replicated DNA. This can be combined with Ki-67 staining, used to distinguish between proliferative and quiescent cells, to report the number of cells in each phase~\citep{eminaga_detection_2016, miller_ki67_2018}. Staining compounds that do not require the cells to be fixed prior to experiments can also be used, allowing for longitudinal time-series tracking of population-level cell cycle phase progression \cite{smith_nuclear_2018, bradford_novel_2006, kimAssayingCellCycle2015}. However, these live-cell dyes can invoke more stress on the cell population, and the resulting measurements have a lower resolution than when cells are fixed \cite{smith_nuclear_2018}.

Mathematical modelling of tumour growth and the cell cycle is an active field of research, spanning a variety of approaches,  including agent-based models, partial differential equations (PDE) and ordinary differential equations (ODE). A review of mathematical modelling of the cell cycle and its impact on treatment strategies is provided by~\cite{maComprehensiveReviewComputational2024}. Several of these \cite{smith_cells_1973, falcoQuantifyingCellCycle2025, celoraDNAstructuredMathematicalModel2022, altinok_automaton_2011, burns_existence_1970, nixson_characterising_2025}, model cell cycle progression via constant transition rates between the different phases in an ODE model. Implicitly, this assumes that the underlying cell cycle phase length follows an exponential distribution. Experimental data shows that cell populations with initial synchronisation in one cell cycle phase will eventually desynchronise over several complete cycles, giving rise to balanced exponential growth (BEG). This growth regime is characterised by an exponentially growing population count, but with cell cycle phase proportions that settle to a time-independent steady state. Flow cytometry measurements of these proportions can then be used to identify the constant transition rates in ODE models of the cell cycle.

Over the past 15 years, Fluorescent Ubiquitination-based Cell Cycle Indicator (FUCCI) has become a widely-used technique to analyse cell cycle progression \cite{zielke_fucci_2015, sakaue-sawanoVisualizingSpatiotemporalDynamics2008}. Rather than staining for DNA content, fluorescent markers are used to indicate accumulation and degradation of cell cycle-regulated proteins. Unlike flow cytometry approaches, FUCCI allows for live tracking of single cells, thus providing higher resolution data such as individual phase durations, lineage tracking and heterogeneity in phase length between cells. An early limitation of FUCCI was the inability to distinguish between $S/G_2/M$ phases, which would all fluoresce green, in contrast to the red of $G_1$ cells. However, reporting of all four phases is now possible using FUCCI4 \cite{bajar_fluorescent_2016}, which uses replication and mitosis-coupled cell cycle markers to distinguish $S$ and $M$ cells, respectively.

Age-structured models of the cell cycle have previously been fitted to both FUCCI imaging and BrdU pulse-labelling data. Weber et al. \cite{weberQuantifyingLengthVariance2014} consider phase lengths following delayed exponential distributions and find that under standard single-pulse BrdU experiments, the means of the phase lengths could be identified, but the variance remained unidentifiable. They propose a re-designed dual-pulse experiment to resolve this issue, highlighting a clear dependence of parameter identifiability on experimental design.

The rise of FUCCI and its higher-resolution derivatives has enabled the investigation of parameter identifiability for more complex underlying phase length distributions. Unlike in the work of Weber et al. \cite{weberQuantifyingLengthVariance2014}, we are no longer limited to devising ways to untangle the population-level data generated by flow cytometry methods, and can instead attempt to leverage the single cell data from FUCCI experiments. For example, Billy et al. \cite{billy_age-structured_2013} use FUCCI time-series data to infer phase-transition distributions. These are then embedded in a PDE system and used to compute population growth rates under different experimental conditions. Such approaches focus on parameter estimation from the experimental data, rather than considering structural or practical identifiability of the parameters for the given experimental data. In another pulse-labelling approach, Sherer et al. \cite{shererIdentificationAgestructuredModels2008} use an experimental design based on extracting a transient subpopulation to infer age-dependent transition rates between cell cycle phases. 

Together, this early work demonstrates that such age-structured models can be parametrised using sufficiently rich experimental data, whether that be BrdU labelling, or FUCCI imaging. Most importantly, each of these three prior works uses time-series data for fitting their model parameters. However, mathematical modellers often do not have access to detailed experimental datasets, instead relying on summary population measurements found in the literature, often from a range of cell lines when complete data on the desired cell line is unavailable. As such, when these rich datasets are not available, it remains unclear which model parameters are identifiable from the most commonly available data types, and how the availability of data influences the ability to identify both model parameters and other biologically meaningful quantities. Unlike previous approaches, which demonstrate that richer experimental designs can recover additional structure, we focus on the complementary problem of parameter inference under constrained summary data.

In this work, we consider an age-structured population model of the cell cycle, in which each phase has an associated PDE, and transition to the next phase depends on the time spent in the current phase. We assume that each phase length follows a delayed-gamma probability distribution, and derive analytical expressions for key measurable quantities. These include the BEG proportions and coefficients of variation of phase lengths, each of which can be experimentally obtained from either flow cytometry or FUCCI, but are often published without the full time-series dataset from which they are derived. By assuming an increasing number of these quantities are available to us, we consider parameter identifiability of our model in different scenarios. Thus, by requiring a given level of detail from the model, we can evaluate the number of these population summary values required to obtain sufficient data for parameter identifiability.

\section{Introducing an age-structured model}
We split the cell cycle into three compartments, corresponding to cells in $G_1$, $S$ and $G_2/M$. An age-structure variable is introduced to track the length of time spent in each phase.

Defining $G_1(a,t)$, $S(b,t)$ and $G_2(c,t)$ to be the densities of cells that have been in the $G_1/S/G_2$ phase for $a$, $b$ and $c$ hours, respectively, we propose the following system of PDEs to describe their evolution:

\begin{equation}
    \frac{\partial G_1}{\partial t} + \frac{\partial G_1}{\partial a} = -\mu_1(a)G_1(a,t) + \beta_4 Q(a,t) - \beta_5 G_1(a,t) , \label{X_pde}
\end{equation}

\begin{equation}
    \frac{\partial S}{\partial t} + \frac{\partial S}{\partial b} = -\mu_2(b)S(b,t), \label{Y_pde}
\end{equation}

\begin{equation}
    \frac{\partial G_2}{\partial t} + \frac{\partial G_2}{\partial c} = -\mu_3(c)G_2(c,t), \label{Z_pde}
\end{equation}

\begin{equation}
    \frac{\partial Q}{\partial t} = -\beta_4Q(a,t) + \beta_5 G_1(a,t), \label{Q_pde}
\end{equation}
where the total number of cells at time $t>0$, $P(t)$, is given by
\begin{equation}
    P(t) = \int_0^\infty \left(G_1(a,t) + Q(a,t)\right) \,da \hspace{0.1cm} + \int_0^\infty S(b,t) \,db \hspace{0.1cm} + \int_0^\infty G_2(c,t) \,dc, \label{P_def}
\end{equation}
and $Q(a,t)$ is the density of quiescent cells that exited $G_1$ with ``age" $a>0$. Here, $\mu_{1}(a), \mu_2(b)$ and $\mu_3(c)$ are the rates at which $G_1/S/G_2$ cells with ages $a$, $b$ and $c$ progress to the next phase of the cell cycle. We note that there is no $a$-derivative in the PDE for $Q(a,t)$, as cell cycle progression is paused during quiescence, and only resumes when cells return to $G_1$. Such cells resume the cell cycle with the same $G_1$ age they had when they became quiescent.

We close equations \eqref{X_pde}-\eqref{Q_pde} by imposing the following boundary conditions:
\begin{equation}
    G_1(0,t) = 2\int_0^\infty\mu_3(c)G_2(c,t) \,dc \hspace{1cm} \text{(cell division),} \label{cell_divis}
\end{equation}
\begin{equation}
    S(0,t) = \int_0^\infty\mu_1(a)G_1(a,t) \,da \hspace{1cm} \text{(progression from $G_1$ to $S$),} \label{progress_bc}
\end{equation}
\begin{equation}
    G_2(0,t) = \int_0^\infty\mu_2(b)S(b,t) \,db \hspace{1cm} \text{(progression from $S$ to $G_2/M$)}. \label{progress_bc2}
\end{equation}

Many healthy human cells experience contact inhibition, which slows cell cycle progression and the population growth rate as the density of cells increase. However, the work presented here is based on experimental data from cancer cells, which typically lose the contact inhibition that prevents normal cell populations from continuing to grow exponentially \cite{pavelContactInhibitionControls2018}. As such, we refrain from imposing any density dependence in the PDE transition rates.

In order to capture cell heterogeneity in intermitotic times (the time taken to progress through the cell cycle) of individual cells, we assume that the rates of cell cycle progression and cell division depend on the time spent in the corresponding compartment. The time a cell spends in each cell cycle phase has been shown to differ between cell lines, and can be described by various probability distributions, including gamma, exponential and log-normal distributions~\citep{yatesMultistageRepresentationCell2017, golubevApplicationsImplicationsExponentially2016, chaoEvidenceThatHuman2019, weberQuantifyingLengthVariance2014,  shererIdentificationAgestructuredModels2008}.

Here we use a delayed gamma distribution as it is analytically tractable. More specifically, we assume that cells in the $G_1$ phase move to the $S$ phase with a probability that follows a shifted gamma distribution in $a>0$, with further shifted gamma distributions representing the transitions from $S$ to $G_2$, and $G_2$ to $G_1$. We include the initial shift to ensure that cells spend at least a fixed, minimum amount of time in each phase, preventing cells jumping from $G_2$ to $S$, for example.

Thus, we assume that cells must spend a minimum time $T_1>0$ in the $G_1$ phase, after which their progression probability to $S$ follows a Gamma$(\alpha_1, \beta_1)$ distribution. The parameter sets $(T_2, \alpha_2, \beta_2$) and $(T_3, \alpha_3, \beta_3)$ define the progression rates for $S$ and $G_2$, respectively. We then use the general formula for deriving age-structured transition rates from probability distributions (presented by Stukalin ~\citep{stukalinAgedependentStochasticModels2013}, Chaffey~\citep{chaffeyEffectG1Transition2014} and Sherer~\citep{shererIdentificationAgestructuredModels2008}) to obtain the functional forms:

\begin{equation}
    \mu_i(a) =
    \begin{cases}
        \frac{\beta_i^{\alpha_i} (a-T_i)^{\alpha_i - 1} \exp(-\beta_i(a - T_i))}{\Gamma(\alpha_i, \beta_i(a-T_i))}, & a > T_i \\
        0, & a < T_i \label{mu_general_def}
    \end{cases}
\end{equation}
where 
\begin{equation}
    \Gamma(\alpha, x) = \int_x^\infty s^{\alpha-1} \exp(-s) ds \label{upper_inc_gam}
\end{equation}
is the upper-incomplete gamma function.
With these functional forms, cells cannot progress to the next cell cycle phase unless they have spent at least a time $T_i$ in their current phase, and their rate of progression asymptotes to a constant value as the time they have spent in that phase increases.

\section{Balanced Exponential Growth (BEG)} \label{BEG}
We expect that an untreated cancer cell population in a nutrient-rich environment with no density restrictions will grow exponentially. Experimental evidence suggests that the proportion of the total cell count in a given cell cycle phase will eventually settle to a constant steady-state as the population grows~\citep{milottiStatisticalApproachAnalysis2008, chiorinoDesynchronizationRateCell2001, olofssonStochasticModelCell2010, vittadelloMathematicalModelsIncorporating2019, gavagninSynchronizedOscillationsGrowing2021}. Thus, the population will eventually settle to a balanced exponential growth (BEG) regime, in which the total cell count grows exponentially, but the proportion of cells in each cell cycle phase becomes constant. As the proportion of cells in a given cell cycle phase can be measured via flow cytometry~\citep{kimAssayingCellCycle2015, pozarowskiAnalysisCellCycle2004}, we will exploit BEG when fitting the PDE model to experimental data.

Under BEG, equations \eqref{X_pde}-\eqref{progress_bc2} admit separable solutions of the form $X(a,t) = \tilde{X}(a) \exp(\lambda t)$ for $X = G_1, S, G_2, Q$, and the PDE model reduces to a system of first-order ODEs. We solve for each $\tilde{X}(a)$ and integrate over the semi-infinite domain of phase ages to obtain
\begin{equation}
    X^{\text{tot}} = \int_0^\infty \tilde{X}(a) \,da, \hspace{0.8cm} \text{for} \hspace{0.3cm} X = G_1, S, G_2, Q.
\end{equation}
By writing $P^{\text{tot}} = G_1^{\text{tot}} + S^{\text{tot}} + G_2^{\text{tot}} + Q^{\text{tot}} $, we calculate the steady-state BEG proportions, $(\bar{G}_1, \bar{S}, \bar{G}_2, \bar{Q})$ as
\begin{equation}
\bar{G}_1 =\frac{G_1^{\text{tot}}}{P^{\text{tot}}}
   = \frac{2\lambda}{\lambda+f(\lambda)} \left(1 - F(\lambda) \right) ,\label{X_prop}
\end{equation}

\begin{equation}
    \bar{S} = 2 \left(1 - \left(\frac{\beta_2}{\beta_2+\lambda}\right)^{\alpha_2} \exp(-\lambda T_2) \right) F(\lambda),  \label{Y_prop}
\end{equation}

\begin{equation}
    \bar{G}_2 = \left(\frac{\beta_3}{\beta_3 + \lambda} \right)^{-\alpha_3}\exp(\lambda T_3) - 1, \label{Z_prop}
\end{equation}

\begin{equation}
    \bar{Q} = \frac{2f(\lambda)}{\lambda+f(\lambda)} \left(1- F(\lambda)\right), \label{Q_prop}
\end{equation}
where 
\begin{equation}
    f(\lambda) = \frac{\lambda\beta_5}{\lambda+\beta_4} \label{f_def}
\end{equation}
and 
\begin{equation}
    F(\lambda) = \exp(-(\lambda+f(\lambda))T_1) \left( \frac{1}{1 +   \frac{\lambda + f(\lambda)}{\beta_1} }\right)^{\alpha_1}. \label{F_def}
\end{equation}

The characteristic equation

\begin{align}
    1 = 2 F(\lambda) \left(\frac{\beta_2}{\beta_2 + \lambda}\right)^{\alpha_2} \left(\frac{\beta_3}{\beta_3 + \lambda}\right)^{\alpha_3} \exp(-(T_2+T_3)\lambda),\label{BEG_char}
\end{align}
defines the unique exponential growth rate, $\lambda$, and enforces that the phase proportions sum to 1.

\section{Fitting to Experimental Data} \label{fitting}
We now consider how we can combine PI and BrdU data with the BEG analysis above to estimate model parameters. We use flow cytrometry data from the RKO cell line (derived from a colon carcinoma) presented by Celora et al.~\citep{celoraDNAstructuredMathematicalModel2022}. These data consist of nine sets of measurements of phase proportions, each measured after the cells have settled into a BEG regime under normoxic conditions ($21\%$ O$_2$). These data do not contain any quiescence proportion measurements, only $G_1$, $S$ and $G_2/M$ proportions and the experimental procedure does not attempt to distinguish between $G_1$ and quiescence. As such, the quiescent proportion is a hidden variable, and so our assumed quiescent proportion must be taken from the measured $G_1$ proportion. Here, we assume a quiescent proportion of $0.061$ of the total cell count, following the experimental results of Corvaisier et. al~\citep{corvaisierRegulationCellularQuiescence2016} for RKO cells. We average over the nine repeats, and present the full set of assumed BEG proportion values in Table \ref{beg_prop_table}.

\begin{table}[h!]
\centering
\begin{tabular}{clc}
\toprule
 & \textbf{Description} & \textbf{Value} \\
\midrule
$\bar{g}_1$& BEG proportion of cells in $G_1$ phase.& 0.246\\
$\bar{s}$& BEG proportion of cells in $S$ phase.& 0.539\\
$\bar{g}_2$& BEG proportion of cells in $G_2$ phase.& 0.154\\
$\bar{q}$& BEG proportion of cells in $Q$ phase.& 0.061\\
\bottomrule
\end{tabular}
\caption{Average BEG phase proportions, for the 9 experimental repeats in Celora et al. \cite{celoraDNAstructuredMathematicalModel2022}.}
\label{beg_prop_table}
\end{table}

For a given cell line of interest, it is often easier to find these single long-term phase proportion values in the literature than a time-series of phase proportion measurements from pulse-labelling experiments (as considered by Weber et al.~\citep{weberQuantifyingLengthVariance2014}).
Since four BEG proportions will be insufficient to determine all nine model parameters, $(\alpha_i, \beta_i, T_i)$, $i = 1,2,3$ we look to combine BEG proportions with population measurements obtained from live single-cell analysis via FUCCI, assuming that the full FUCCI imaging data are unavailable.

By recording the phase lengths of individual cells using FUCCI techniques, it is possible to obtain data on heterogeneity within the population. Here, we use the coefficient of variation of phase lengths (CV), and the minimum length of each phase,  denoted by $T_i$ in the PDE model \eqref{X_pde}-\eqref{Q_pde}. For a probability distribution with mean $\mu$ and standard deviation $\sigma$, the coefficient of variation is defined by 
\begin{equation}
    \text{CV} = \frac{\sigma}{\mu},  \label{CV definition}
\end{equation}
and so for a shifted gamma distribution, with location parameter $T$, shape parameter $\alpha$ and scale parameter $\beta$, we obtain
\begin{equation}
    \text{CV}_i = \frac{\sqrt{\alpha_i}}{\alpha_i + T_i\beta_i} \hspace{1cm} i = 1,2,3. \label{CV gamma}
\end{equation}

From equations \eqref{X_prop} and \eqref{Q_prop}, we find that
\begin{equation}
    f(\lambda) = \frac{\bar{Q}}{\bar{G_1}} \lambda. \label{f_alt_def}
\end{equation}
As we have values for $\bar{G}_1$ and $\bar{Q}$, and $\beta_4$, $\beta_5$ appear only via $f(\lambda)$ in equations \eqref{X_prop}-\eqref{Q_prop}, we postpone consideration of $\beta_4$ and $\beta_5$ for the remainder of this work.

With $T_i$, $i = 1,2,3$ known, and consideration of $\beta_4$, $\beta_5$ postponed, there remain six unknown parameter values: $\alpha_i$ and $\beta_i$,  $i = 1,2,3$. Equations \eqref{X_prop}-\eqref{Z_prop} and \eqref{CV gamma} define six algebraic equations which we can use to determine the six unknown parameters. 

These expressions lead us to consider the structural and practical identifiability of the model. A model is structurally identifiable if given model output, there is a unique input parameter set~\citep{wielandStructuralPracticalIdentifiability2021, liuParameterIdentifiabilityModel2024, raueStructuralPracticalIdentifiability2009}. In our case, we say that the model is structurally identifiable if each of the six unknown parameters can be uniquely defined from the BEG proportions, CV values and minimum phase lengths.

Practical identifiability refers to the ability to uniquely estimate model parameters from noisy output data~\citep{wielandStructuralPracticalIdentifiability2021, liuParameterIdentifiabilityModel2024, raueStructuralPracticalIdentifiability2009}. This corresponds to being able to accurately identify the true parameter values from noisy measurements of BEG proportions.

With fewer than nine pieces of data it is not possible to uniquely determine all nine model parameters, and so the model is structurally unidentifiable. However, we still consider parameter groupings that remain identifiable even with limited available data.

\subsection{Case 1: Only BEG data are available}
Firstly, we assume that we have no access to values derived from single-cell data, leaving only the mean phase proportions in the BEG regime from flow cytometry. This corresponds to the simplest, cheapest and quickest experimental setup, and so these values are often the easiest to find in the literature. We also note that the growth rate $\lambda$ is related to the doubling time, $T_D$, of the cell line via $\lambda = \log(2)/T_D$. Henceforth, we will assume a doubling time $T_D = 22$ hours, consistent with experimental estimates of 21-23 hours for RKO cells~\citep{hailekaColonCancerCells2019, witzelAnalysisImpedancebasedCellular2015}.

Since the model is structurally unidentifiable in this case, we consider parameter groupings that are identifiable from such data. With only three pieces of data available, we expect a maximum of three identifiable parameter groupings. In this case, we are fitting for all nine model parameters, so $\vec{x} = (\alpha_1, \alpha_2, \alpha_3, \beta_1, \beta_2, \beta_3, T_1, T_2, T_3)$. In fact, the identifiable parameter groupings can be seen from the analytical expressions for the phase proportions in equations \eqref{X_prop}-\eqref{Q_prop}, and are given by 
\begin{equation}
    \eta_1(\alpha_1, \beta_1, \gamma_1) = \left(\frac{\beta_1}{\beta_1 + \lambda(1 + \bar{Q}/\bar{G}_1)}\right)^{\alpha_1} \exp(-\lambda (1 + \bar{Q}/\bar{G}_1)T_1) = 1 - \frac{\bar{Q} + \bar{G}_1}{2}, \label{eta_1}
\end{equation}
\begin{equation}
    \eta_2(\alpha_2, \beta_2, \gamma_2) = \left(\frac{\beta_2}{\beta_2 + \lambda}\right)^{\alpha_2} \exp(-\lambda T_2) = \frac{1 + \bar{G}_2}{2 - \bar{Q} - \bar{G}_1}, \label{eta_2}
\end{equation}

\begin{equation}
    \eta_3(\alpha_3, \beta_3, \gamma_3) = \left(\frac{\beta_3}{\beta_3 + \lambda}\right)^{\alpha_3} \exp(-\lambda T_3) = \frac{1}{1+\bar{G}_2}. \label{eta_3}
\end{equation}
For a given set of phase proportions, $(\bar{G}_1, \bar{S}, \bar{G}_2, \bar{Q})$, these groupings restrict the tuples $(\alpha_i, \beta_i, T_i)$, $i = 1,2,3$, to a surface in parameter space. The natural bounds $\alpha_i>1$ and $\beta_i, T_i>0$ do little to constrain the parameter space further. Instead, we turn to look at identifiability of the mean and variance of each phase length across parameter space.

Recalling that the mean and variance of a Gamma$(\alpha, \beta)$ distribution with delay $T$ are given by
\begin{equation}
    \text{Mean} = \frac{\alpha}{\beta} + T, \hspace{1cm}\text{Var =} \frac{\alpha}{\beta^2}, \hspace{1cm}\label{mean_and_var}
\end{equation}
we look at maximising and minimising these biologically-relevant quantities subject to the constraints in equations \eqref{eta_1}-\eqref{eta_3} above, in order to assess how constrained they are under this limited data. We leave the full mathematical details to the Supplementary Materials, noting that the results align with those of Weber et al. \cite{weberQuantifyingLengthVariance2014} in the simplified delayed-exponential case $\alpha = 1$, and focus attention on the $G_1$ phase only, thus making use of equation \eqref{eta_1} and considering $(\alpha_1, \beta_1, T_1)$. The other phases can be analysed in an identical way.

\begin{figure}
    \centering
    \includegraphics[width=1\linewidth]{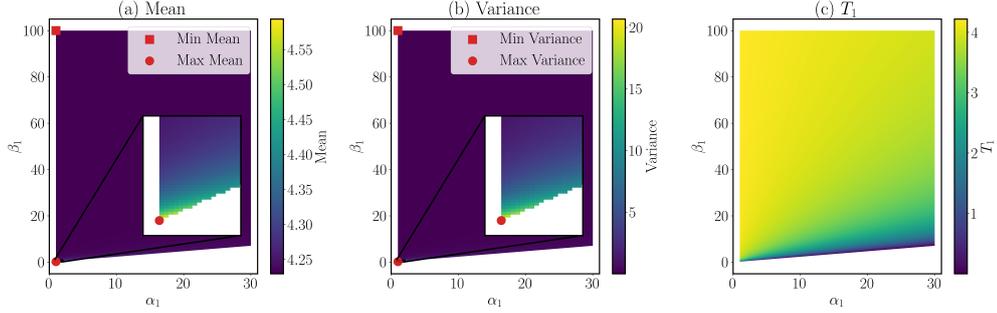}
    \caption{Heatmaps showing the values of the (a) mean, (b) variance and (c) minimum phase delay $T_1$ of the $G_1$ phase length demonstrate small variance in mean value, and larger discrepancies in variance and $T_1$ across $(\alpha_1, \beta_1)$ parameter space. The value of $T_1$ is calculated from equation \eqref{eta_1} using BEG proportions in Table \eqref{beg_prop_table},  which is then used to calculate the mean and variance from equations \eqref{mean_and_var}. We exclude unfeasible regions where $T_1<0$.}
    \label{fig:mean_var}
\end{figure}
Figure \ref{fig:mean_var} shows how the mean, variance and $T_1$ values of the shifted gamma distribution for the $G_1$ phase length vary across $(\alpha_1, \beta_1)$ parameter space subject to the relation in equation \eqref{eta_1} under the BEG proportions from Table \eqref{beg_prop_table}. Whilst most of the displayed parameter space fixes both the mean and variance to be at the lower end of the total range, we find a small region near $\alpha_1=1$ and small $\beta_1$ in which the mean and variance reach their maximum. However, the mean values are constrained to be within 0.4 hours of each other across parameter space, and so account for at most around a 10$\%$ perturbation from the minimum value. 

On the other hand, the variance takes a much larger range of values across parameter space, with the higher values concentrated in regions where the mean is also near its maximum. This is consistent with the results of Weber et al.~\citep{weberQuantifyingLengthVariance2014}, which show that the mean of their delayed-exponential distribution is much more narrowly constrained by just BEG proportions than the variance.

The maximum and minimum occur at the same position in parameter space for both the mean and variance. The impact of this can be seen when looking at the $T_1$ values at these two points. When the mean and variance are minimised, $T_1$ is much larger than when the mean and variance are maximised. Furthermore, the value of $T_1$ at the minimum mean is close to the value of this minimum mean. This means that most of the contribution towards the mean in expression \eqref{mean_and_var} comes from $T_1$, the fixed delay, and hence explains why the variance of the $G_1$ phase length in this case is small. On the other hand, at maximum mean, which is also relatively close to the minimum mean, $T_1$ has a much lower contribution to the mean. Thus, as a higher contribution to the mean comes from the random Gamma$(\alpha_1, \beta_1)$ distribution than the fixed delay, we see why the variance in $G_1$ phase length is also much higher in this case.

While varying $(\alpha_1, \beta_1)$ across valid parameter space makes little impact on the mean $G_1$ phase progression, the results above suggest that parameter choice will impact underlying PDE dynamics. The PDE system in equations \eqref{X_pde}-\eqref{Q_pde} will settle into a BEG regime from any initial distribution after sufficient time, but the distribution of $G_1$ phase lengths affects the time taken to reach this steady state behaviour. Smaller variance in $G_1$ phase lengths due to a large contribution to the mean from $T_1$ suggests that all cells entering $G_1$ at a given time will show more synchronicity in their $G_1$ exit times, and therefore more synchronicity throughout the cell cycle, than when $T_1$ contributes less to the mean phase length.

\begin{figure}
    \centering
    \includegraphics[width=1\linewidth]{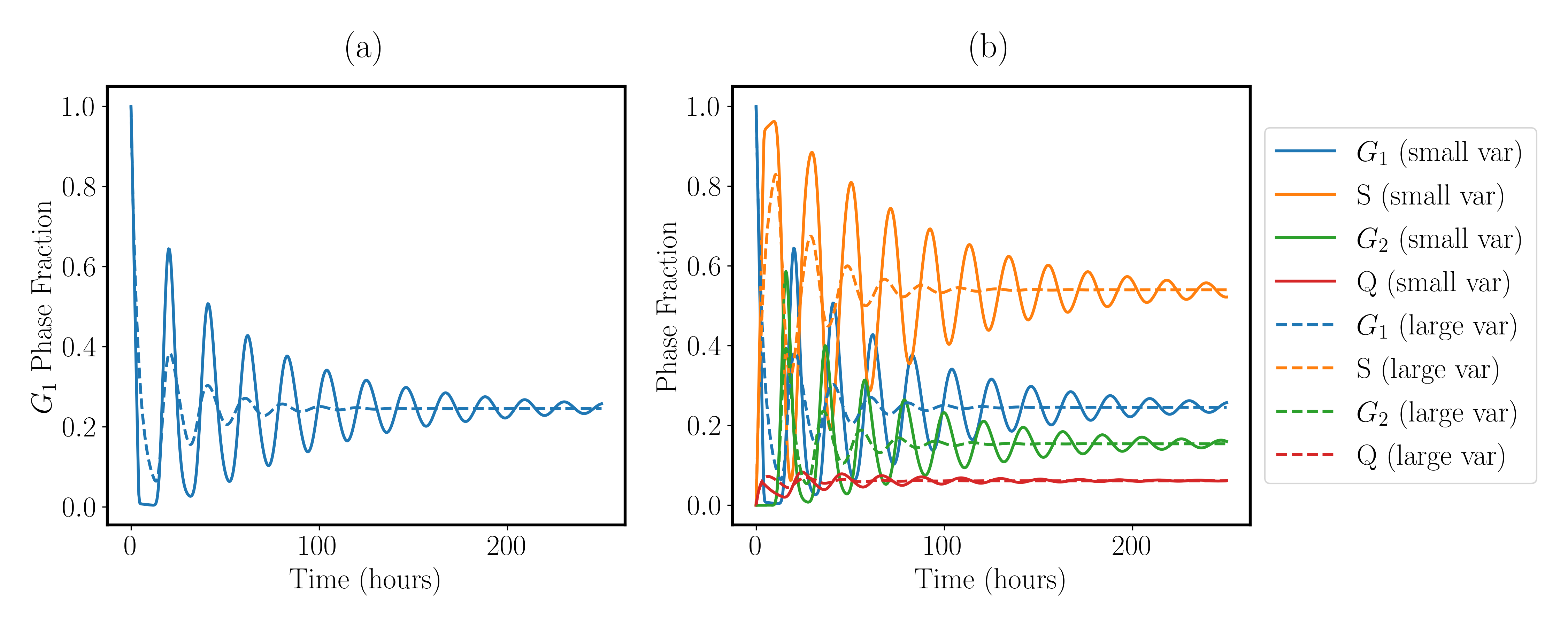}
    \caption{Simulations of the full PDE system \eqref{X_pde}-\eqref{Q_pde} show that less variance in the values of the $G_1$ phase lengths delays the return to BEG significantly from cases of higher variance. In both cases, $(\alpha_i, \beta_i, T_i)$, $i = 2,3$ are listed in Table \eqref{unique_fit_table}. (a) The $G_1$ phase proportion over time for the two different variance cases. (b) All four phase proportions over time for the two different variance cases. For the smaller variance case, $(\alpha_1, \beta_1, T_1) = (1, 5, 4.03)$, whilst for the larger variance case,
    $(\alpha_1, \beta_1, T_1) = (1, 0.22, 0.046)$. Both simulations had identical initial conditions, which had all cells starting in $G_1$.}
    \label{fig:delayed_desync}
\end{figure}

This is illustrated in Figure \ref{fig:delayed_desync}, where we simulate the full PDE system \eqref{X_pde}-\eqref{Q_pde} with all cells initially in $G_1$ for the two extreme cases of small and large variance in the $G_1$ phase lengths. As expected, we see that a smaller variance in $G_1$ phase length causes the system to take longer to reach BEG than in the larger variance case. Moreover, the amplitude of the intermediate oscillations in phase proportions is much higher in the small variance case.

This work considers parametrising a structured-PDE system, which we could then use to simulate responses to different anti-tumour treatments and look at how the underlying cell cycle dynamics drive those responses. As these treatments are likely to perturb cell cycle dynamics away from BEG by specifically targeting cells in a given phase, Figure \ref{fig:delayed_desync} illustrates the dangers of making arbitrary parameter choices from the $(\alpha, \beta)$ parameter space in Figure \ref{fig:mean_var}, despite the lack of impact on mean $G_1$ phase lengths. Once the treatment is applied and the system is perturbed, the choice of phase length parameters impacts how long the system takes to return to the steady state. Figure \ref{fig:delayed_desync} shows that this delay in returning to BEG can differ from around 3 days to 10 days, which could impact the predictive performance and accuracy of the model under fractionated treatment spaced a few days apart.

\begin{figure}
    \centering
    \includegraphics[width=0.75\linewidth]{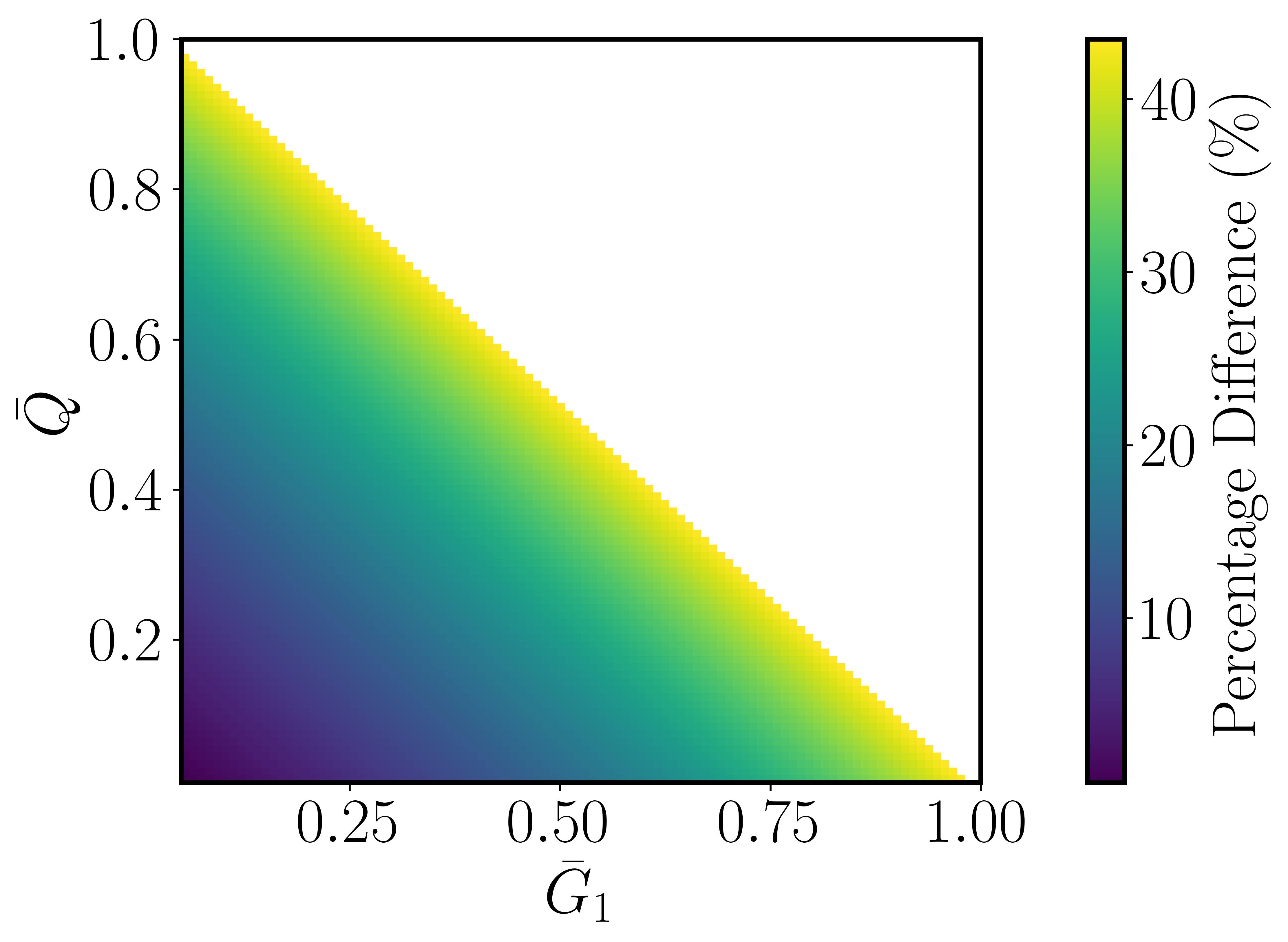}
    \caption{The percentage difference between the max and minimum mean $G_1$ phase lengths captured by the PDE decreases as the BEG $\bar{G}_1$ and $\bar{Q}$ proportions vary. See main text for the definition of percentage difference.}
    \label{fig:prop_difference}
\end{figure}

As mentioned above, for the BEG proportions in Table \ref{beg_prop_table}, Figure \ref{fig:mean_var} suggests that the mean length of the $G_1$ phase is determined to within 0.4 hours, a relatively small difference when the minimum mean is around 4.2 hours. We now consider how robust this result is to variations in the BEG proportions. For each $\bar{G}_1$ and $\bar{Q}$ in feasible parameter space ($\bar{G}_1 + \bar{Q} < 1$), we use equations \eqref{eta_1} and \eqref{mean_and_var} to calculate the minimum and maximum mean and variance across $(\alpha_1, \beta_1)$ space, as done above for the single BEG proportion set in Table \eqref{beg_prop_table}. Figure \ref{fig:prop_difference} shows the percentage difference in the range of means, calculated by obtaining the difference between the maximum and minimum mean values as a percentage of the minimum, across feasible phase proportions. We see that the relative range of means is lower for smaller combinations $(\bar{G}_1, \bar{Q})$. Therefore, if we only seek to use the structured PDE model to identify the mean phase lengths, rather than pin down individual phase duration distribution parameters, our estimates for the $G_1$ length will be more precise for cell lines with smaller experimental values of $\bar{G}_1$ and $\bar{Q}$, demonstrating the suitability of the model for certain cell lines over others.

\subsection{Case 2: Available data includes BEG proportions and cell phase length CV values} \label{CV only}

We now consider the case where we have access to some population-level measurements derived from FUCCI data. Given that FUCCI imaging records the phase lengths of individual cells, a relevant population measurement for our model \eqref{X_pde}-\eqref{Q_pde} would be the minimum observed length of each phase. However, an accurate estimate of the minimum phase lengths from single-cell tracking experiments requires frequent imaging over the course of the experiment. Depending on the resolution needed, imaging on the order of every 10 minutes or less may be necessary to capture minimum phase lengths. In this case, we assume that imaging was not frequent enough to identify minimum phase lengths. 

Instead, we consider a single population metric that is not as impacted by imaging resolution as the minimum phase length. A natural choice might be the mean of each phase length. However, as we saw in Case 1, knowing only the BEG proportions is enough to identify a small range of possible mean values. Taking the $G_1$ analysis above and repeating for the $S$ and $G_2$ phases, we find that the mean phase lengths are restricted to $[4.23, 4.60]$, $[12.17, 14.84]$ and $[4.55, 4.89]$ hours, respectively, for the BEG proportions in Table \ref{beg_prop_table}. If we were to include FUCCI estimates of mean phase lengths in our subsequent analysis, they would need to fall within these intervals for the analysis to be viable.

Without access to FUCCI cell cycle data for RKO cells or even population-level summaries, we instead search for single-cell data from similarly characterised cell lines to accompany BEG proportions in Table \ref{beg_prop_table}. RKO cells are characterised by a relatively long $S$-phase~\citep{celoraDNAstructuredMathematicalModel2022}, and data from Chao et al.~\citep{chaoEvidenceThatHuman2019} identifies the U2OS cell line as also having a long $S$ phase (mean $\approx$ 10 hours and $G_1$ and $G_2$ phases of approximately 4 hours). However, the mean $S$ phase length falls outside of the BEG-derived interval of $[12.17, 14.84]$ hours for RKO cells, and so mean phase lengths of U2OS cells are not suitable candidates for the following analysis. In a scenario where the cell line of choice had FUCCI-derived mean phase lengths values available, which also agree with the BEG-derived mean phase length intervals, then the subsequent analysis could be repeated by fitting to the mean phase length values.

The lack of available FUCCI summary data for RKO cells justifies our use of the coefficient of variation of phase lengths rather than mean values; across cell lines, the $S$ phase has the least variability in phase duration \cite{weiner_inferring_2024, cameron_evidence_1963}, and so from the definition in equation \eqref{CV definition}, the $S$ phase has the smallest CV value. The relative order of the coefficient of variation for $G_1$ and $G_2$ phase lengths varies across cell lines, but are both larger than that of $S$. Given that this work intends to act as a proof of concept, we therefore use the single-cell phase length data from \cite{chaoEvidenceThatHuman2019} for U2OS cells to estimate the coefficients of variation of the phase lengths, which are given in Table \ref{COV_table}.

\begin{table}[ht!]
\centering
\begin{tabular}{clc}
\toprule
 & \textbf{Description} & \textbf{Value} \\
\midrule
$CV_1$& CoV of $G_1$ phase lengths.& 0.42 (h)\\
$CV_2$& CoV of $S$ phase lengths.& 0.13 (h)\\
$CV_3$& CoV of $G_2$ phase lengths.& 0.33 (h)\\

\end{tabular}
\caption{Parameter values for the coefficients of variation, which we calculated from single-cell data in Chao et al.~\citep{chaoEvidenceThatHuman2019} for U2OS cells.}
\label{COV_table}
\end{table}

In this case, we are now fitting for six parameters, so $\vec{x} = (\alpha_1, \alpha_2, \alpha_3, T_1, T_2, T_3)$, with $\beta_i$ ($i = 1,2,3$) given by
\begin{equation}
    \beta_i = \frac{\sqrt{\alpha_i}(1 - \text{CV}_i\sqrt{\alpha_i})}{\text{CV}_i \times T_i}. \label{beta_def}
\end{equation}
We also leverage the BEG proportion data as in Case 1 to identify three parameter groupings, given by equations \eqref{eta_1}-\eqref{eta_3}, which specify $T_i$ as functions of $\alpha_i$ and $\beta_i$. Focusing again on the $G_1$ phase, these two algebraic relations for each $(\alpha_1, \beta_1, T_1)$ generate a curve of parameters satisfying the natural constraints $\alpha_1>1$, $\beta_1 > 0$, $T_1 > 0$ (full mathematical analysis can be found in the Supplementary Materials). By considering $T_1 = T_1(\beta_1)$ and $\alpha_1 = \alpha(\beta_1)$ along this curve, we again look at the mean and variance of the $G_1$ phase lengths.

\begin{figure}
    \centering
    \includegraphics[width=0.75\linewidth]{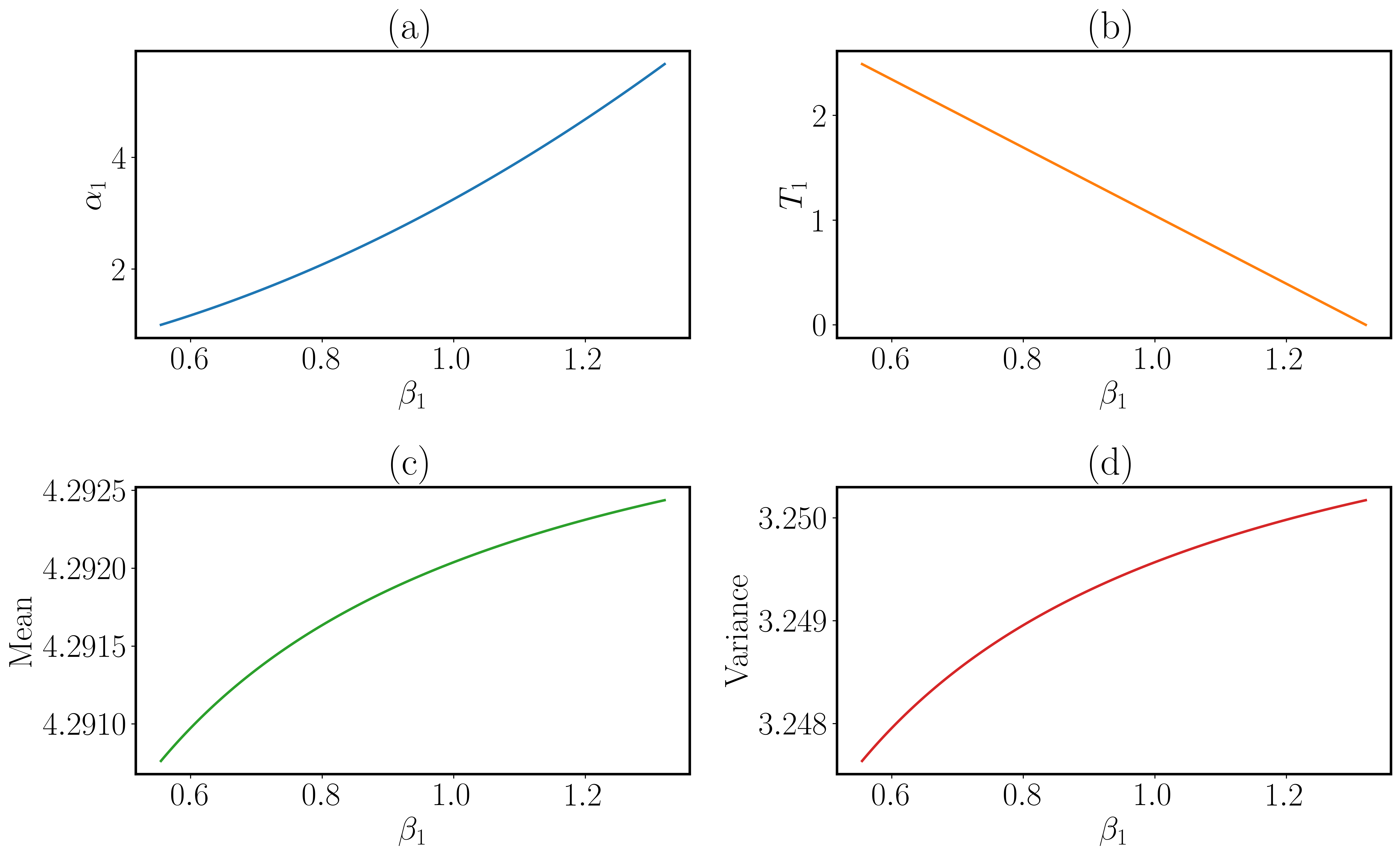}
    \caption{The mean and variance of the $G_1$ phase lengths remain consistent across feasible parameter space, despite variations in $\alpha_1$, $\beta_1$ and $T_1$ when the CV and BEG values are fixed. (a) $\alpha_1(\beta_1)$, (b) $T_1(\beta_1)$, (c,d) mean and variance of $G_1$ phase length calculated via \eqref{mean_and_var}, across all viable $\beta_1$ values. }
    \label{fig:avmv}
\end{figure}
Figure \ref{fig:avmv}(a) and (b) show that $\alpha_1$ and $T_1$ vary considerably across feasible values of $\beta_1$ for the BEG proportions and CV values in Tables \ref{beg_prop_table} and \ref{COV_table}. However, panels (c) and (d) show that the mean can be specified to a precision of 0.002 hours, and variance to within 0.03. FUCCI imaging designed to investigate cell cycle phase lengths typically takes images at 0.1-0.2 hour intervals. Thus, even with instantaneous phase transition times (the time taken for a cell to transition from definitely being categorised in one phase, to definitely being categorised in the next), the precision obtained by these experiments is worse than our model estimate.

In conclusion, temporally-coarse FUCCI imaging that prohibits estimates of the minimum phase lengths, but permits estimates of the CV values and BEG proportions allows us to specify the first two moments of the underlying gamma distributions, despite the lack of structural identifiability in the model. Therefore, if estimates of these two moments are all that is required from the model, the summary data from temporally-coarse FUCCI imaging is sufficient.

\subsection{Case 3: Available data includes BEG proportions, cell phase length CV values and minimum phase lengths} \label{fit_1}
Finally, we consider the case in which FUCCI summary data from temporally finer (images taken every $0.1-0.2$ hours) imaging are available, corresponding to experimental imaging that was sufficiently frequent to specify minimum phase lengths. We again use the U2OS single cell data in Chao et al. \cite{chaoEvidenceThatHuman2019}, and the values are given in Table \ref{T_table}.

\begin{table}[ht!]
\centering
\begin{tabular}{clc}
\toprule
 & \textbf{Description} & \textbf{Value} \\
$T_1$& Minimum phase length of $G_1$.& 1.8 (h)\\
$T_2$& Minimum phase length of $S$.& 7.7 (h)\\
$T_3$& Minimum phase length of $G_2$.& 1.7 (h)\\
\end{tabular}
\caption{Parameter values for the minimum phase lengths, which we calculated from the single-cell data in Chao et al.~\citep{chaoEvidenceThatHuman2019} for U2OS cells.}
\label{T_table}
\end{table}

\subsubsection{Uniqueness of fit}
Using the estimates for the BEG phase proportions, minimum phase lengths and CV values in Tables \ref{beg_prop_table}, \ref{COV_table} and \ref{T_table}, our optimisation problem is to minimise
\begin{equation}
    n(\vec{x}) = (\bar{G}_1(\vec{x}) - \bar{g}_1)^2 + (\bar{S}(\vec{x}) -\bar{s})^2 + (\bar{G}_2(\vec{x}) - \bar{g}_2)^2, \label{min_prob}
\end{equation}
where $\vec{x} = (\alpha_1, \alpha_2, \alpha_3, \beta_1, \beta_2, \beta_3)$, the six unknown parameters. We use SciPy's \texttt{differential$\_$evolution} function to perform the minimisation. This method searches large regions of parameter space, and can be parallelised for computational efficiency \cite{storn_differential_1997}.

The algorithm need only fit for $\alpha_i$, $i = 1,2,3$, as $\beta_i$ can be calculated via equation \eqref{beta_def}. We constrain $1 < \alpha_i < 1/CV_i ^2$ for $i = 1,2,3$ so that $\beta_i$ and the coefficients $\mu_i(a)$ in equations \eqref{beta_def} and \eqref{mu_general_def} are positive and finite, respectively.

We run the minimisation algorithm 100 times to confirm that there is a unique parameter set satisfying our minimisation problem, given in Table \ref{unique_fit_table}. This unique parameter set has $n(\vec{x}) < 10^{-30}$, where $n(\vec{x})$ is given by equation \eqref{min_prob}. Since $T_i$ are prescribed by the values in Table \ref{T_table}, and $\beta_i$ are defined by equation \eqref{beta_def}, which guarantees that the delayed gamma distributions will have the desired CV values for any given $\alpha_i$, the value of $n(\vec{x})$ measures how close each BEG phase proportion calculated from the expressions \eqref{X_prop}-\eqref{Z_prop} is to data given in Table \ref{beg_prop_table}. Each squared term in equation \eqref{min_prob} is the difference between one analytical BEG phase proportion, and the corresponding data value, both of which must be bounded between 0 and 1. Thus, $n(\vec{x}) < 10^{-30}$ indicates excellent agreement between each analytical phase proportion, and the BEG data.
\begin{table}[h!]
\centering
\begin{tabular}{clc}
\toprule
 & \textbf{Description} & \textbf{Value} \\
\midrule
$\alpha_1$& Shape parameter of $G_1$ gamma distribution.& 1.92\\
$\alpha_2$& Shape parameter of $S$ gamma distribution.& 8.08\\
$\alpha_3$& Shape parameter of $G_2$ gamma distribution.& 3.63\\
$\beta_1$& Rate parameter of $G_1$ gamma distribution.& 0.78 \\
$\beta_2$& Rate parameter of $S$ gamma distribution.& 1.79\\
$\beta_3$& Rate parameter of $G_2$ gamma distribution.& 1.26\\

\bottomrule
\end{tabular}
\caption{The unique sets of gamma distribution parameters obtained when we fit the model to the BEG phase proportions in Table \ref{beg_prop_table}, and CV and minimum phase lengths in Tables \ref{COV_table} and \ref{T_table}.}
\label{unique_fit_table}
\end{table}

\subsubsection{Minimum phase length impacts quality of fit}
As shown above, the values of $T_i$ in Table \ref{T_table} are capable of producing an excellent fit of model parameters to BEG data. We now consider how probing the $T_i$ parameter space impacts the maximum quality of fit when minimising $n(\vec{x})$ in equation \eqref{min_prob}.

In order to visualise how $T_i$ affects the quality of model fit, we consider a reduced model by merging the $S$ and $G_2$ phases to remove the need for the parameters $(\alpha_3, \beta_3, T_3)$ in the model. In doing so, the system becomes
\begin{equation}
    \frac{\partial G_1}{\partial t} + \frac{\partial G_1}{\partial a} = -\mu_1(a)G_1(a,t) + \beta_4(P(t)) Q(a,t) - \beta_5(P(t)) G_1(a,t) , 
\end{equation}

\begin{equation}
    \frac{\partial S}{\partial t} + \frac{\partial S}{\partial b} = -\mu_2(b)S(b,t),
\end{equation}

\begin{equation}
    \frac{\partial Q}{\partial t} = -\beta_4(P(t))Q(a,t) + \beta_5(P(t)) G_1(a,t),
\end{equation}
with
\begin{equation}
    G_1(0,t) = 2\int_0^\infty\mu_2(b)S_2(b,t) \,db, \hspace{1cm} S(0,t) = \int_0^\infty\mu_1(a)G_1(a,t) \,da. 
\end{equation}
For this reduced model there are three BEG proportions, and we aim to estimate $(\alpha_i, \beta_i)$ for $i = 1,2$ only.

\begin{figure}
    \centering
    \includegraphics[width=1\linewidth]{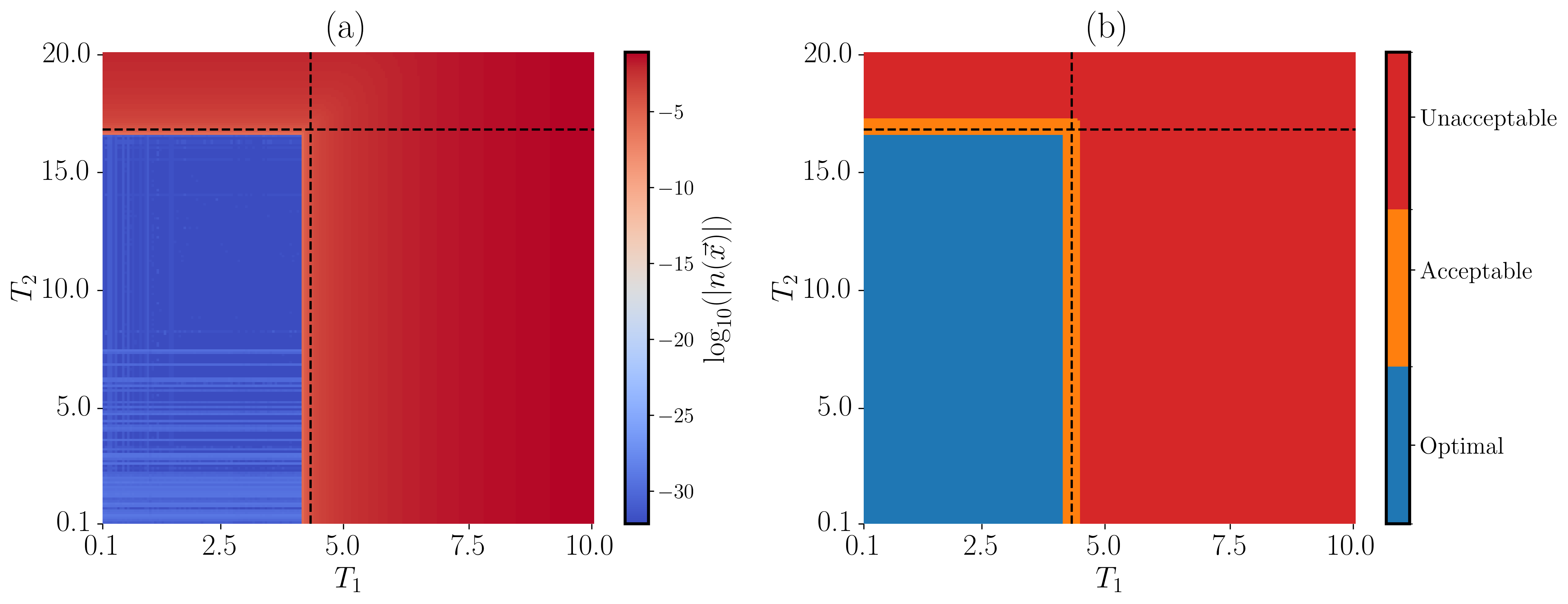}

    \caption{Varying the imposed minimum phase lengths $(T_1, T_2)$ in a reduced two-phase cell cycle model identifies a rectangular region in parameter space that produces an excellent unique fit to the prescribed BEG phase proportions. (a) A heatmap of the log-transformed value of $n(\vec{x})$ at the fitted parameters, and (b) classification of each point in the parameter space as producing an ``unacceptable", ``acceptable" or ``optimal" fit to the BEG proportions, as defined in the main text. The black dashed lines represent the analytical bounds $T_1^*$ and $T_2^*$, defined by equations \eqref{T1_ub} and \eqref{T2_T3_ub}.}
    \label{fig:unique_fit_regions}
\end{figure}

In Figure \ref{fig:unique_fit_regions}, we fix the coefficients of variation for the $G_1$ and $S/G_2/M$ phase lengths to be 0.42 and 0.13, respectively, with BEG proportions 0.247 and 0.693, respectively. We show how the goodness of fit changes when we fit for $(\alpha_i, \beta_i)$ ($i = 1,2$) when $T_1$ and $T_2$ vary. For all $(T_1, T_2)$ pairs, the differential evolution algorithm converges to a global minimum. However, inspection of the values of $n(\vec{x})$ for these parameter sets indicates that the fit is poor for most of the parameter space. More specifically, we define an ``unacceptable" fit to have $n(\vec{x}) > 0.0004$ when evaluated at the fitted parameter set. If we allow $n(\vec{x}) = 0.0004$, then the BEG phase proportions could differ from the data by up to 0.02. This definition is largely arbitrary, but justified by noting that a 0.02 difference from the quiescence BEG proportion of 0.061 is a $33\%$ error. This definition can be altered based on the required BEG proportions. Here, we consider any value of $n(\vec{x}) \in [10^{-6}, 0.0004]$ to be an acceptable fit, with the lower bound corresponding to a maximum difference between data and fitted BEG proportions of 0.001, and any value below this to be an optimal fit.

In fact, Figure \ref{fig:unique_fit_regions} suggests that there is only a small boundary between minimum phase lengths with an unacceptable fit to the data, and those that allow for an optimal fit (taken to be $n(\vec{x}) < 10^{-6}$). These simulations show that the rectangular region of $(T_1, T_2)$ space that allows for this optimal fit is approximately $(T_1, T_2) \in ([0, 4.13], [0, 16.60])$.  Hence, we see a well-defined region of $(T_1, T_2)$ parameter space that produces an optimal unique fit to the data. 

Furthermore, as in Weber et al.~\citep{weberQuantifyingLengthVariance2014}, we can derive an analytical upper bound for $T_i$ in the four-phase model. For example, by rearranging equation \eqref{X_prop} for $\beta_1$ and noting that the resulting expression is decreasing in $T_1$ with an asymptote at $T_1 = T_1^* > 0$, we find that
\begin{equation}
    T_1 < T_1^* =- \frac{1}{\lambda + f(\lambda)} \log\left(\frac{2 - \bar{G}_1 - \bar{Q}}{2}\right). \label{T1_ub}
\end{equation}

Similarly, we can obtain the following bounds for $T_2$ and $T_3$:
\begin{equation}
    T_2 < T_2^* =  -\frac{1}{\lambda} \log\left(\frac{2 - \bar{G}_1 - \bar{Q} - \bar{S}}{2 - \bar{G}_1 - \bar{Q}}\right), \hspace{0.5cm} T_3 < T_3^* =  \frac{1}{\lambda} \log(1 + \bar{G}_2). \label{T2_T3_ub}
\end{equation}
In the three-phase model considered in Figure \eqref{fig:unique_fit_regions}, we can set $\bar{G}_2 = 0$ and obtain $T_1 < 4.25$ and $T_2 < 16.72$. We see that the region of optimal fit is contained within these bounds.

In conclusion, if we assume that the coefficient of variation of each phase length is known, along with experimentally-obtained BEG fraction, we are restricted in the values of the minimum cell cycle phase lengths, $T_i$, that we can impose in order to obtain an optimal parameter fit for the model \eqref{X_pde}-\eqref{Q_pde}. From equations \eqref{T1_ub} and \eqref{T2_T3_ub}, we can derive upper bounds for $T_i$, based on the known BEG fractions. However, the results in Figure \ref{fig:unique_fit_regions} suggest that in order to get an optimal fit to the data, the values of $T_i$ must be restricted further. If data values of $T_i$ are not within this restricted region, there is a choice to be made as to whether to prioritise fitting to BEG proportions exactly (using $T_i$ values that correspond to low $n(\vec{x})$), or to allow BEG proportions to stray from their measured values in order to use exact data values of $T_i$.

\subsubsection{Fitting to noisy biological data} \label{noisy_data}
The results above suggest that a unique best-fit parameter set can be identified by fixing the CV values and the minimum phase lengths, with excellent agreement under the condition that minimum phase lengths are chosen appropriately. We conclude that, in this case, the model is structurally identifiable \cite{wielandStructuralPracticalIdentifiability2021}. We now consider noisy data to check for practical identifiability \cite{wielandStructuralPracticalIdentifiability2021}. Here, the noisy data is the nine repeats for BEG proportions of RKO cells from Celora et al.~\cite{celoraDNAstructuredMathematicalModel2022}.

Due to the lack of quiescent proportions in the experimental data from Celora et al.~\citep{celoraDNAstructuredMathematicalModel2022} we first consider simulating these in an appropriate manner. Given that data was calculated by dividing the number of cells in each phase by the total number of cells in the population, a sensible candidate for the underlying distribution of cells across the four phases would be a binomial model \cite{mccullaghGeneralizedLinearModels2019}. This would capture the discrete possible combinations of proportions when the total number of cells is fixed. However, due to the absence of the total number of cells in the data, we use the Dirichlet distribution~\citep{albert_denis_2012_dirichlet_jags}. This allows us to directly sample proportion sets, as it hardwires the fact that the proportions must sum to one. The resulting \textit{in silico} data set is provided in Table \ref{tab:full_data}, and full details of utilising a Dirichlet distribution to simulate quiescent proportions can be found in the Supplementary Material.

\begin{table}
    \centering
    \begin{tabular}{c|c|c|c}
    \multicolumn{1}{c|}{$G_1$} & \multicolumn{1}{c|}{$S$} &
\multicolumn{1}{c|}{$G_2$}&
\multicolumn{1}{c}{$Q$} \\
\hline
0.2556 & 0.5570 & 0.1518 & 0.0356 \\
0.2305 & 0.5338 & 0.1685 & 0.0672 \\
0.2214 & 0.6021 & 0.1216 & 0.0549 \\
0.3103 & 0.4757 & 0.1324 & 0.0817 \\
0.1920 & 0.5400 & 0.1900 & 0.0780 \\
0.2675 & 0.4985 & 0.1476 & 0.0864 \\
0.2465 & 0.5157 & 0.1879 & 0.0499 \\
0.2386 & 0.5714 & 0.1391 & 0.0510 \\
0.2480 & 0.5607 & 0.1475 & 0.0438 \\
    \end{tabular}
    \caption{Cell cycle phase proportions, based on those in Celora et. al \cite{celoraDNAstructuredMathematicalModel2022}, and extended to account for the quiescent compartment.}
    \label{tab:full_data}
\end{table}
We use a Bayesian inference approach to assess the practical identifiability of the model for this extended in-silico dataset, full details of which can be found in the Supplementary Materials.

\begin{figure}[h!]
    \centering
    \includegraphics[width=0.8\linewidth]{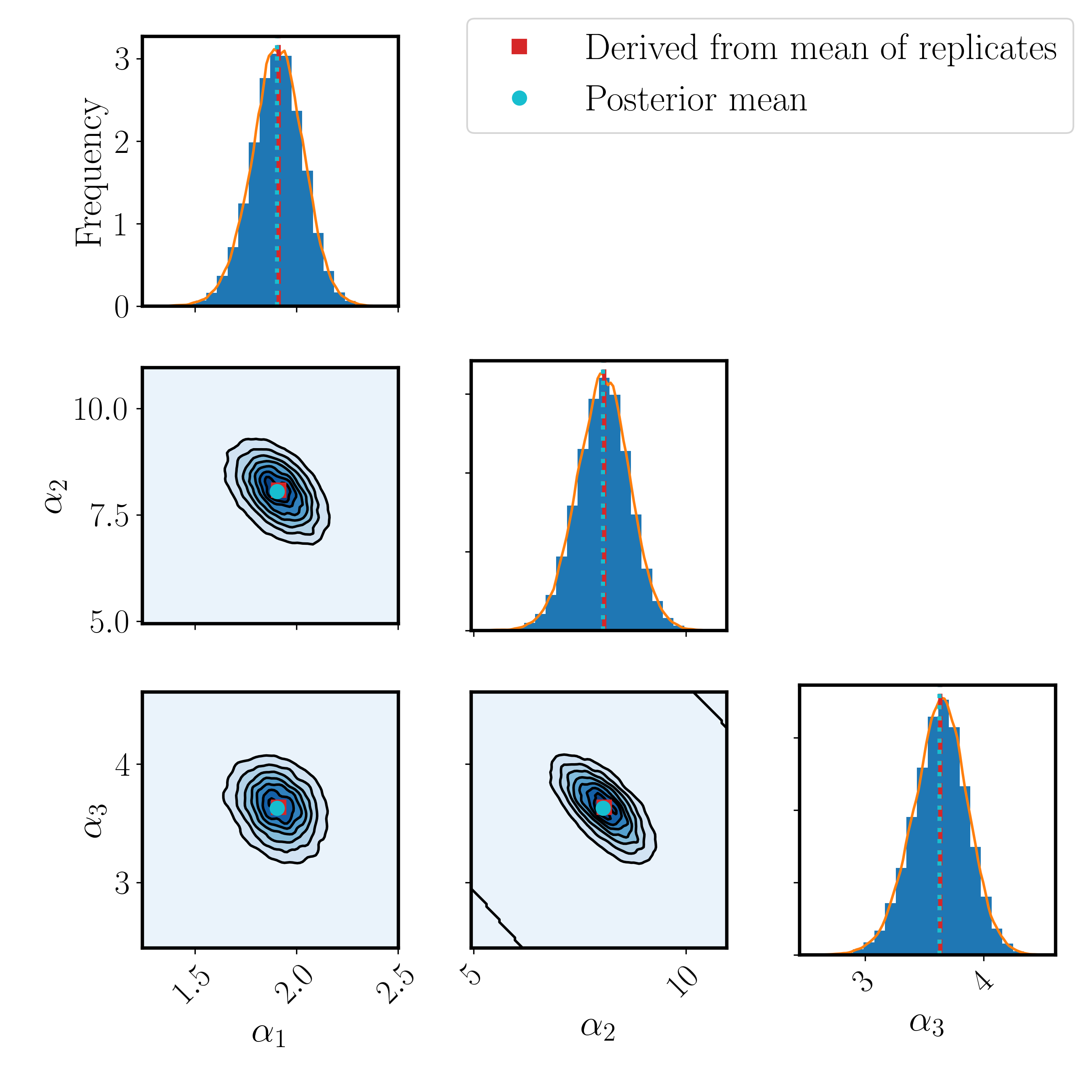}
    \caption{Marginal posterior distributions for the parameters $\alpha_1, \alpha_2, \alpha_3$ demonstrate practical identifiability in each case. For each distribution along the diagonal, the red dashed line corresponds to the parameter value found by minimising Equation \eqref{min_prob} where the input proportions are the mean values of each column in Table \eqref{tab:full_data}. The blue dotted line represents the mean value of the marginal posterior distribution. The off-diagonal plots show contours of the joint distributions of each $\alpha_i$ pair.}
    \label{fig:marginal_posteriors}
\end{figure}

We estimate the posterior distributions of the unknown parameters $(\alpha_1, \alpha_2, \alpha_3)$ using a Markov Chain Monte Carlo (MCMC) method via the Python package PINTS (Probabilistic Inference on Noisy Time-Series) \cite{clerxProbabilisticInferenceNoisy2019}. The resulting marginal posterior distributions in Figure \ref{fig:marginal_posteriors} indicate that each parameter is symmetrically distributed about a single, well-defined peak. We conclude that these parameters are practically identifiable from the data in Table \ref{tab:full_data} \cite{dalyInferencebasedAssessmentParameter2018, browningIdentifiabilityAnalysisStochastic2020}. Furthermore, the mean value of each marginal posterior distribution aligns well with the parameter value found by minimising equation \eqref{min_prob} with the average cell cycle phase proportions of the data found in Table \ref{beg_prop_table}, i.e. the pointwise estimates of $\alpha_i$ found by minimising the sum of squared errors. This suggests that, for these data, taking the mean of the BEG data in Table \eqref{tab:full_data} and fitting the $\alpha_i$ ($i = 1,2,3$) by minimising $n(\vec{x})$ in equation \eqref{min_prob} is equivalent to the means of the posterior distributions of $\alpha_i$ found via a full MCMC analysis.

We conduct a further check of practical identifiability of the parameters $(\alpha_1, \alpha_2, \alpha_3)$ using profile likelihood analysis for each one, with the results presented in Figure \ref{fig:PLA} \cite{kreutzProfileLikelihoodSystems2013} (full details can be found in the Supplementary Materials). From this analysis, we can estimate confidence intervals for the parameter values using the likelihood ratio statistic \cite{wielandStructuralPracticalIdentifiability2021, kreutzProfileLikelihoodSystems2013}. In Figure \ref{fig:PLA}, the horizontal line indicates the 95\% confidence interval about the maximum likelihood estimate for each parameter, and is defined by
\begin{equation}
    \text{CL}_{95}(\alpha_i) = \{\alpha_i : |\text{PL}(\alpha_i) - \text{PL}(\hat{\alpha}_i)| < \chi^2_{1, 0.95}\}, \label{CI_calc}
\end{equation}
where $\hat{\alpha}_i= \operatorname*{arg\,max}_{\alpha_i \in [0, 1/CV_i^2]} \text{PL}(\alpha_i)$ is the maximum likelihood estimate of $\alpha_i$, where $\text{PL}(\alpha_i)$ is the profile likelihood function of $\alpha_i$. The parameter $\alpha_i$ is said to be practically identifiable if the interval $\text{CL}_{95}(\alpha_i)$ is finite. We find that this is indeed the case for each $\alpha_i$, $i \in \{1,2,3\}$ and so each parameter is practically identifiable from the data in Table \ref{tab:full_data}.

\begin{figure}[h!]
    \centering
    \includegraphics[width=1\linewidth]{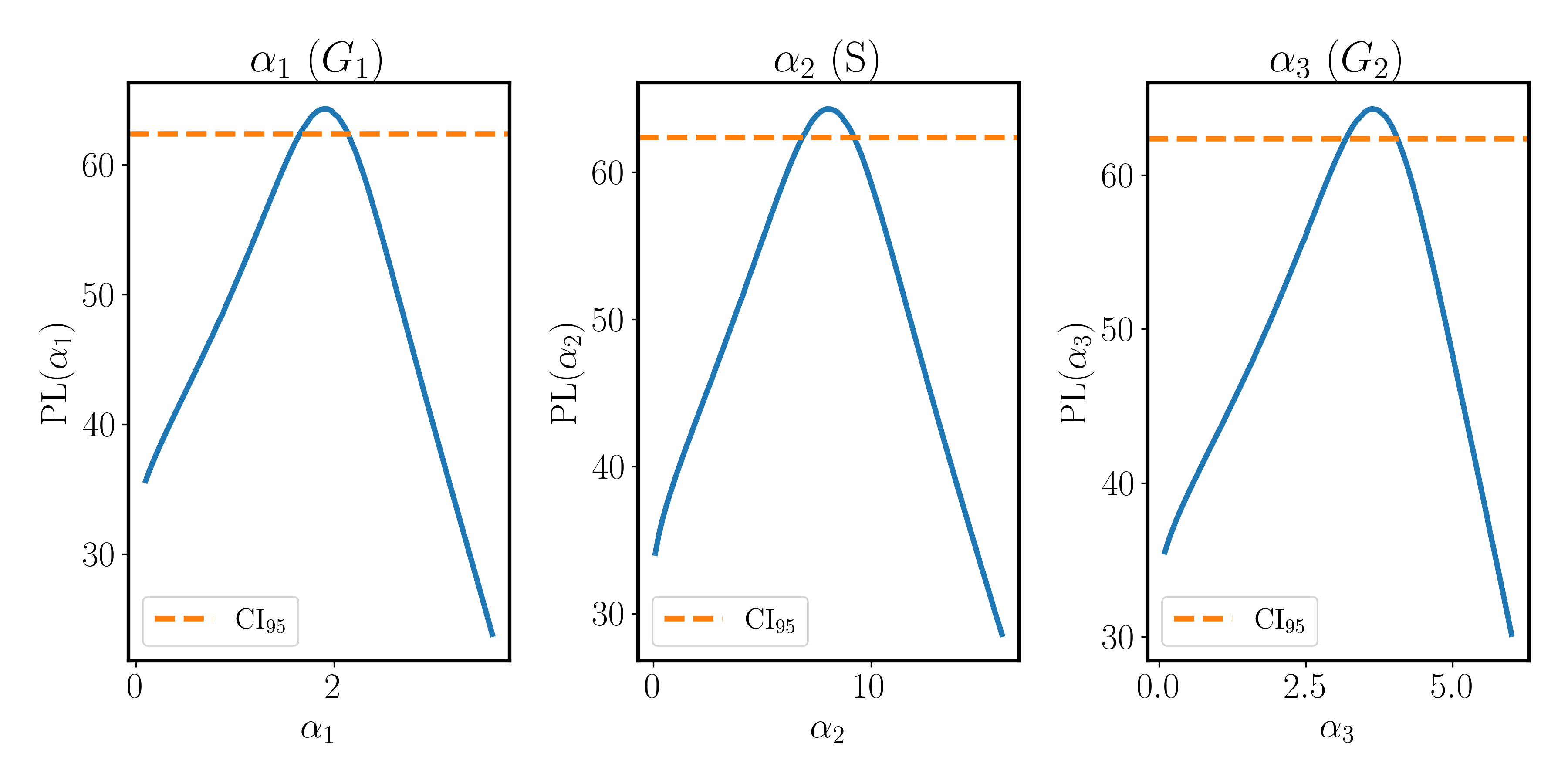}
    \caption{Profile likelihood analysis for each of the model parameters, $\alpha_1$, $\alpha_2$ and $\alpha_3$, also indicates practical identifiability. Each plot displays the profile likelihood for the model parameters $\alpha_1, \alpha_2$ and $\alpha_3$ as a blue curve. The intersections of the dashed orange line with the profile likelihood represent the bounds of the $95\%$ confidence interval defined by equation \eqref{CI_calc}.  }
    \label{fig:PLA}
\end{figure}

\section{Discussion}

Our analysis demonstrates that the availability of population-averaged data for parametrising mathematical models of the cell cycle strongly influences model identifiability. However, we find that limited available data still permits identifiability of key parameter groupings, corresponding to the first moments of the phase-length distributions.

The common lack of available time-series data for cell cycle experiments across cell lines means mathematical modellers often must collate summary data from multiple sources to parametrise models. Since cell cycle data arises mainly from 
population-level (DNA-stained flow cytometry), and single-cell phase tracking (FUCCI) experiments, a suitable mathematical model frames cell cycle dynamics in terms of phase durations. Our age-structured PDE formulation with delayed gamma-distributed phase lengths allows for analytical tractability, and provides sufficient detail to integrate population-level, and single-cell measurements. Furthermore, balanced exponential growth regimes arise naturally from this formulation, which is important for mirroring experimental observations in cell populations \cite{vittadelloMathematicalModelsIncorporating2019, nowakImpactVariabilityCell2023}, and also produces a natural way to link the model to the simpler flow cytometry data, which is often easier to obtain.

This study highlights that structural unidentifiability due to lack of appropriate data does not preclude the usefulness of the model. For cases in which the amount of data prohibits unique parameter estimations, we are still able to recover estimates of biologically meaningful quantities, such as the mean and variance of each cell cycle phase duration. We thus conclude that moments of these phase-residence time distributions are more robustly identifiable than individual distribution parameters.

Furthermore, even in the case where the model appears structurally identifiable in the sense that a unique parameter set minimises the objective function, the quality of the parameter fit is determined by the agreement between the different data. In particular, BEG proportions obtained from flow cytometry are sufficient to identify a relatively small range of possible average phase lengths. As such, these provide a limiting bound on the single-cell minimum phase length data that can produce a good fit. When these data are inconsistent, this framework based on patchwork data naturally produces a poorer fit. Our framework instead identifies that using a measure that transfers more consistently across cell lines, such as coefficient of variation, can provide a way around this issue.

This study illustrates the importance of considering the intended purpose of the model when looking for experimental data. If estimating the mean cell cycle phase length is the goal, then obtaining only BEG proportions is sufficient to estimate these values under the framework presented here. However, one must be careful when choosing individual model parameters for further use, such as simulating treatment. Large variations in the variances of the phase lengths across parameter space lead to different transient dynamics as the cells settle back into BEG growth, impacting predictions when fractionated treatment is simulated.

When additional information such as phase length variance or individual model parameters are needed, more detailed experimental data are required. We used the coefficient of variation of phase length and minimum phase length, both of which were calculated from single-cell FUCCI tracking data, which are more complex, expensive and time-consuming to collect experimentally, and often not publicly available in the literature. As such, we conclude that there is a trade-off between the ease with which necessary data can be uncovered, and the amount of information we can obtain about the underlying processes via this age-structured PDE model. 

The analysis assumed that density-dependent cell cycle progression effects could be ignored due to the lack of contact inhibition in the cancer cell populations that provide the data. Therefore, further work might aim to extend our analysis to account for normal cells, in which density dependence plays a vital role in cell cycle regulation. However, this would likely reduce the analytical tractability when deriving expressions for the BEG phase proportions, increasing the computational cost required to repeat the analysis. If suitable data was to become available, we would also like to repeat the practical identifiability analysis presented in Section \ref{noisy_data} for the case where experimentally-measured quiescent proportions were available, along with single-cell analysis, for a single cell line.

Of course, given well resolved time-series data as in Weber et al., Billy et al., and Sherer et al.~\cite{weberQuantifyingLengthVariance2014, billy_age-structured_2013, shererIdentificationAgestructuredModels2008}, we would turn to the methods used within those works to parametrise our age-dependent phase transition rates. However, the framework presented here is driven by data limitations, rather than methodological advances for full data sets. By sensibly building up the amount of population summary data assumed to be available in each case, our framework shows that collating data across cell lines and experimental set-ups can be effective in parametrising an age-structured PDE model of the cell cycle.

\clearpage

\appendix

\section{Supplementary Information}

\subsection{Optimising mean and variance from identifiable parameter groupings}
\subsubsection{Case 1: Only BEG data is available}
As noted in the main text, inspection of Equations (12)-(17) allows us to find three identifiable parameter groups, namely

\begin{equation}
    \eta_1(\alpha_1, \beta_1, T_1) = \left(\frac{\beta_1}{\beta_1 + L}\right)^{\alpha_1} \exp(-L T_1) = 1 - \frac{\bar{Q} + \bar{G}_1}{2}, \label{eta_1_SI}
\end{equation}
\begin{equation}
    \eta_2(\alpha_2, \beta_2, T_2) = \left(\frac{\beta_2}{\beta_2 + \lambda}\right)^{\alpha_2} \exp(-\lambda T_2) = \frac{1 + \bar{G}_2}{2 - \bar{Q} - \bar{G}_1}, \label{eta_2_SI}
\end{equation}

\begin{equation}
    \eta_3(\alpha_3, \beta_3, T_3) = \left(\frac{\beta_3}{\beta_3 + \lambda}\right)^{\alpha_3} \exp(-\lambda T_3) = \frac{1}{1+\bar{G}_2}, \label{eta_3_SI}
\end{equation}
where $L = \lambda(1 + \bar{Q}/\bar{G}_1)$. We now wish to consider the parameter values $(\alpha_i, \beta_i, T_i)$, $i = 1,2,3$, that minimise and maximise the values of the Gamma distribution means and variances, subject to $\eta_i(\alpha_i, \beta_i, T_i) = \eta_i$, a constant value.

For brevity, we present the method for the $G_1$ parameters $(\alpha_1, \beta_1, T_1)$ only, noting that the other parameter sets $(\alpha_i, \beta_i, T_i)$, $i = 2,3$, can be analysed identically.

For fixed, known $\bar{Q}$ and $\bar{G}_1$, we can rearrange equation \eqref{eta_1_SI} to obtain

\begin{equation}
    T_1 = \frac{\alpha_1 \log{(\frac{\beta_1}{\beta_1 + L})} - \gamma_1}{L}, \label{T1_def}
\end{equation}
where $\gamma_1 = \log(1 - (\bar{Q} + \bar{G}_1)/2) < 0$.

Using the fact that the mean of a delayed Gamma distribution with shape parameter $\alpha$, rate parameter $\beta$ and delay $T$ is $\alpha/\beta + T$,  we define
\begin{equation}
    m(\alpha_1, \beta_1) = \alpha_1 \left(\frac{1}{\beta_1} + \frac{\log({\frac{\beta_1}{\beta_1+L}})}{L} \right)  - \frac{\gamma_1}{L}. \label{M_def}
\end{equation}

Thus, our problem is reduced to minimising or maximising $m(\alpha_1, \beta_1)$ over $\beta_1 > 0$ and $\alpha_1>1$. Another natural boundary can be derived from imposing $T_1 > 0$, which forces
\begin{equation}
    \alpha_1 < \frac{\gamma_1}{\log({\frac{\beta_1}{\beta_1+L}})}, \label{alpha_UB}
\end{equation}
and, combining this with $\alpha_1>1$, we also obtain
\begin{equation}
    \beta_1 > \frac{L}{\exp(-\gamma_1) - 1}. \label{beta_UB}
\end{equation}
Calculation of the partial derivatives shows that $m(\alpha_1, \beta_1)$ is increasing in $\alpha_1$ and decreasing in $\beta_1$. Consolidating all of this, we find that
\begin{equation}
    \min{m(\alpha_1, \beta_1)} = -\frac{\gamma_1}{L} \hspace{1cm} \text{at} \hspace{0.5cm} \alpha_1 = 1, \beta_1 = \infty, \label{max_mean}
\end{equation}
and 
\begin{equation}
    \max m(\alpha_1, \beta_1) = \frac{\exp(-\gamma_1)-1}{L} \hspace{1cm} \text{at} \hspace{0.5cm} \alpha_1 = 1, \beta_1 = \frac{L}{\exp(-\gamma_1)-1},
\end{equation}
allowing us to calculate the range of the minimum and maximum, as presented in Figure 3 of the main text. We note that these minima and maxima expressions agree with those of Weber et al. \cite{weberQuantifyingLengthVariance2014}, who consider a delayed exponential model (equivalent to $\alpha = 1$ in our delayed-gamma formulation).

\subsubsection{Case 2: Available data includes BEG proportions and CV values}
When we also assume knowledge of the CV values for each phase length, our feasible parameter space for $(\alpha_1, \beta_1)$ shrinks. Using the definition of the coefficient of variation of a shifted gamma distribution, we find that
\begin{equation}
    T_1 = \frac{\sqrt{\alpha_1} (1 - CV_1 \sqrt{\alpha_1})}{CV_1 \beta_1}.
\end{equation}
Pairing this with equation \eqref{T1_def}, we find the following two possible values of $\alpha_1$ for a given $\beta_1$:
\begin{equation}
\alpha_1^{\pm}(\beta_1)= \left( \frac{\frac{1}{\beta_1 CV_1} \pm \sqrt{\frac{1}{(CV_1 \beta_1)^2} + \frac{4\gamma_1}{L}\left(\frac{H(\beta_1)}{L} + \frac{1}{\beta_1}\right)}}{2\left(\frac{H(\beta_1)}{L}+ \frac{1}{\beta_1}\right)} \right) ^2,
    \label{alpha_PM}
\end{equation}
where $H(\beta_1) = \log(\beta_1/(\beta_1+L))$.

As in Case 1, we need $\beta_1 > 0$, $T_1 > 0$ and $\alpha_1 > 1$. The constraint $\beta_1 > 0$ implies that $\alpha_1 < 1/CV_1^2$, which, via some rearranging, suggests that $\alpha_1^+(\beta_1)$ is unfeasible for all $\beta_1>0$.

By noting that $\alpha_1^{-}(\beta_1)$ is increasing in $\beta_1$, our feasible region is constrained to $\beta_1 \in [\hat{\beta}_1, \beta_1 ^*]$, where $\alpha_1^-({\hat{\beta}_1}) = 1 $ and $\alpha_1^-(\beta_1^*) = 1/CV_1^2$. Solving equation \eqref{alpha_PM}, we find that $\hat{\beta}_1$ is the unique positive solution to
\begin{equation}
    1 + \frac{L}{\beta_1} = \exp\left( \frac{L (CV_1 - 1)}{\beta_1 CV_1} -\gamma_1\right),
\end{equation}
and
\begin{equation}
    \beta_1^* = \frac{L}{\exp(-CV_1^2\gamma_1) - 1}.
\end{equation}

It is possible to prove that $T_1>0$ for all  $\beta_1 \in [\hat{\beta}_1, \beta_1 ^*]$.

\subsection{Simulating in-silico quiescent proportions}
As discussed in the main text, the RKO cell cycle phase data does not include quiescent cell proportions. To generate in-silico estimates for these, we assume that each measurement of the proportions $(\bar{G}_1, \bar{S}, \bar{G}_2, \bar{Q})$ is drawn from a Dirichlet distribution \cite{albert_denis_2012_dirichlet_jags}. 

More specifically, for a Dirichlet distribution of dimension $M$ we write
\begin{equation}
    (X_1, X_2, X_3,..., X_M) \sim \text{Dir}(c_1, c_2, c_3,..., c_M),
\end{equation}
where
\begin{equation}
    \mathop{\mathbb{E}}(X_i) = \frac{c_i}{c_0}, \hspace{1cm} \text{Var}(X_i) = \frac{c_i (c_0-c_i)}{c_0^2 ( 1 + c_0)}, \hspace{1cm } \text{for } i=1,2,3,...M, \label{def_dirich}
\end{equation}
and $c_0 = \sum_{i=1}^{i=M} c_i$.

\begin{table}
\begin{tabular}{c|c|c}
    \multicolumn{1}{c|}{$G_1$} & \multicolumn{1}{c|}{$S$} &
\multicolumn{1}{c}{$G_2$} \\
\hline
    0.2912 & 0.5570 & 0.1518 \\
    0.2978 & 0.5338 & 0.1685 \\
    0.2763 & 0.6021 & 0.1216 \\
    0.3919 & 0.4757 & 0.1324 \\
    0.2700 & 0.5400 & 0.1900 \\
    0.3539 & 0.4985 & 0.1476 \\
    0.2964 & 0.5157 & 0.1879 \\
    0.2895 & 0.5714 & 0.1391 \\
    0.2918 & 0.5607 & 0.1475 \\
    \end{tabular}
    \caption{Balanced exponential growth phase proportions, as recorded by Celora et. al \cite{celoraDNAstructuredMathematicalModel2022}.}
    \label{tab:Giulia_data}
\end{table}
    
We start by approximating the parameters for the underlying Dirichlet distribution of the 9 experimental repeats shown in Table \ref{tab:Giulia_data} . We identify the standard deviations of the $S$ and $G_2$ phase proportions to be $\sigma_S = 0.038794$ and $\sigma_{G_2} = 0.023666$, respectively. Experimental results from Corvaisier et. al \cite{corvaisierRegulationCellularQuiescence2016} suggest that the standard deviation for the quiescent phase is approximately $\sigma_Q = 0.018$. We assume that $(\bar{G}_1, \bar{S}, \bar{G}_2, \bar{Q}) \sim \text{Dir}(c_1, c_2, c_3, c_4)$, where $c_1 = \bar{g}_1 c_0$, $c_2 = \bar{s} c_0$, $c_3 = \bar{g}_2 c_0$ and $c_4 = \bar{q} c_0$ ($(\bar{g}_1, \bar{s}, \bar{g}_2, \bar{q})$ are defined in Table 1 of the main text), and $c_0$ is chosen to minimise the difference between the observed and calculated standard deviations in a least-squares sense. More specifically, using equation \eqref{def_dirich} we choose $c_0$ to minimise
\begin{equation}
  \text{Error}_{SD}^2(c_0) =  \left(\sigma_S - \sqrt{\frac{\bar{s}  (1 - \bar{s})}{1 + c_0}}\right)^2 + \left(\sigma_{G_2} - \sqrt{\frac{\bar{g}_2  (1 - \bar{g}_2)}{1 + c_0}}\right)^2 + \left(\sigma_Q - \sqrt{\frac{\bar{q}  (1 - \bar{q})}{1 + c_0}}\right)^2.
\end{equation}

We estimate that $c_0 \approx 141.3$. Using the conditional distribution property for Dirichlet distributions \cite{albert_denis_2012_dirichlet_jags}, we deduce further that 
\begin{equation}
    \frac{1}{s + g_2}(\bar{G}_1, \bar{Q})|(\bar{S} = s, \bar{G}_2 = g_2) \sim  \text{Dir}(c_1, c_4), \label{conditional}
\end{equation}

Thus, given values of $s$ and $g_2$ for each replicate (see Table \ref{tab:Giulia_data}), we can sample from the above distribution to estimate the corresponding $G_1$ and $Q$ proportions. In order to ensure that the nine samples have $\bar{q} = 0.061$ and $\sigma_Q = 0.018$ and that the values are reproducible, we generate 10,000 random seeds before sampling, and choose the one producing samples that minimise
\begin{equation}
    \text{Error}_{\text{mean}}^2 = (\bar{q} - 0.061)^2 + (\sigma_Q - 0.018)^2,
\end{equation}
where $\bar{q}$ and $\sigma_Q$ are the sample mean and standard deviation. This process generates our in-silico dataset in Table 5 in the main text.

\subsection{Bayesian Inference}
We use a Bayesian inference approach to assess practical identifiability of the model to the in-silico data in Table 5 of the main text.

Following Celora et. al \cite{celoraDNAstructuredMathematicalModel2022}, we denote by $F_j^i$ ($j \in (1, 2, 3, 4)$) the measurement in the $i$-th row and $j$-th column of Table 5.  As we assume that the coefficients of variation, minimum phase lengths and doubling times are fixed, our unknown parameter set is $\theta = (\alpha_1, \alpha_2, \alpha_3)$.  We pool the data from Table 5 in the set $\mathcal{E} = \left\{(F_1^i, F_2^i, F_3^i, F_4^i)\right\}_{i = 1}^9$ and denote by $\vec{w}(\theta ) = (\bar{G}_1(\theta ), \bar{S}(\theta ), \bar{G}_2(\theta ), \bar{Q}(\theta ))$ the values of the cell cycle phase proportions for parameter values $\theta $.

Assuming that the data in Table 5 arises from noisy realisations of $\vec{w}(\theta )$ under a Dirichlet distribution model, we also consider an unknown $c_0$ parameter as defined in equation \eqref{def_dirich} in the Bayesian procedure. We assume that $\text{log}(c_0)$ is uniformly distributed in $[\log(2), \log(100000)]$ (recall that $c_0$ controls the variance of the Dirichlet distribution), and use a uniform prior on $[0.1, 1/CV_i]$ for $\alpha_i$ ($i = 1,2,3)$.

The log-likelihood of this problem is then
\begin{equation}
    \ell(\theta , c_0) = 9\log(\Gamma(c_0)) - 9\sum_{j = 1}^4 \log(\Gamma(c_j(\theta ))) + \sum_{i=1}^9 \sum_{j=1}^4 (c_j(\theta )-1)\log(F_j^i), \label{log_pdf}
\end{equation}
where $(c_1(\theta ),c_2(\theta ), c_3(\theta ), c_4(\theta )) = c_0 \vec{w}(\theta ) $.

We implement this prior and log-likelihood in the Python package PINTS (Probabilistic Inference on Noisy Time-Series) \cite{clerxProbabilisticInferenceNoisy2019} and use a Markov Chain Monte Carlo (MCMC) method to estimate the corresponding posterior distribution. To initialise the chains, we use the maximum a posteriori (MAP) estimate of $(\theta ,c_0)$, which we compute by minimising the negative log-posterior arising from log-likelihood \eqref{log_pdf} and the uniform/log-uniform priors. For the MCMC algorithm we run three chains using \texttt{HaarioBardenetACMC} from the PINTS library, with 100,000 iterations and a burn-in set of 5000 iterations.

To assess the results of the MCMC run, we start by computing the Gelman-Rubin statistic, $\hat{R}$, and the effective sample size (ESS) for each parameter in $\theta$ and $c_0$. For each parameter, $\hat{R}$ indicates whether or not the MCMC chains have mixed well by comparing within-chain variance to between-chain variance of the posterior samples, with $\hat{R} \rightarrow 1$ as the number of iterations increases for a convergent run. We compute $\hat{R}$ to be $(1.00022627, 1.00008, 1.00006592, 1.00015525)$ for the parameters $(\alpha_1, \alpha_2, \alpha_3, c_0)$, respectively. These values are below the recommended upper threshold of 1.01, suggesting that the 3 chains mix well. The ESS is also used to evaluate chain mixing. As successive samples from the MCMC algorithm are autocorrelated, the ESS refers to the equivalent number of independent samples that generates the same information. Therefore, a higher ESS represents better chain mixing. We compute ESS values of $(20000, 20908, 21041, 17028)$, once again above the recommended threshold of 400.

For the subsequent profile-likelihood analysis, we generate a uniform grid of points that lie within the support of $\alpha_i$ (recall that the support of $\alpha_i$ is $[0, 1/CV_i^2]$. We fix $\alpha_i$ at each of these values and fit for the remaining parameters $c_0$ and $\alpha_j$, $j \neq i$, by maximising the log-likelihood, $\ell(\theta , c_0)$, in equation \eqref{log_pdf}. For each value of $\alpha_i$, the profile likelihood is defined as
\begin{equation}
\text{PL}(\alpha_i) = \operatorname*{max}_{\theta_{-i}, c_0} \ell(\theta, c_0), \label{PL}
\end{equation}
where $\theta_{-i} = \theta \backslash \{\alpha_i\}$.

\backmatter

\bmhead{Acknowledgements}
RN is supported by the Engineering and Physical Sciences Research Council (Grant No.
EP/W524311/1).

\section*{Declarations}

\subsection*{Funding}
RN is supported by the Engineering and Physical Sciences Research Council (Grant No.
EP/W524311/1). For the purpose of open access, the authors have applied a CC BY public copyright license to any author accepted manuscript arising from this submission.

\subsection*{Competing interests}
The authors have no competing interests to declare that are relevant to the content of this article.

\subsection*{Code availability}
The code used to generate the figures in this publication can be accessed by contacting ruby.nixson@maths.ox.ac.uk.

\subsection*{Author contribution}
All authors contributed to the study conception and design. Formal analysis and investigation were performed by Ruby E. Nixson. The first draft of the manuscript was written by Ruby E. Nixson and all authors commented on previous versions of the manuscript. All authors read and approved the final manuscript.

\bibliography{bibliography}

\end{document}